\documentclass[11pt,aps,prd,superscriptaddress,reprint,preprintnumbers,nofootinbib,floatfix]{revtex4-1}

\usepackage[utf8]{inputenc}
\usepackage{graphicx}
\usepackage{bbold}
\usepackage{comment}
\usepackage[normalem]{ulem}  
\usepackage{xcolor}

\usepackage{subcaption}  
\usepackage{makecell}
\usepackage{afterpage}

\usepackage{amsmath,amssymb,amsfonts}

\usepackage{physics}

\usepackage[breaklinks]{hyperref}
\hypersetup{colorlinks=true,allcolors=[rgb]{0.5,0.0,0.5}}

\newcommand{\sw}{\sin^2\theta_W}

\newcommand{\orcid}[1]{\href{https://orcid.org/#1}{#1}}

\newcommand{\ord}[1]{\mathcal{O}{(#1)}}
\newcommand{\beq}{\begin{equation}}
	\newcommand{\eeq}{\end{equation}}
\newcommand{\bea}{\begin{eqnarray}}
	\newcommand{\eea}{\end{eqnarray}}
\newcommand{\eps}{\varepsilon}

\newcommand{\eMMsc}{\epsilon_{\mu\mu}^\mathrm{iS}}
\newcommand{\eTTsc}{\epsilon_{\tau\tau}^\mathrm{iS}}
\newcommand{\eMMvec}{\epsilon_{\mu\mu}^\mathrm{iV}}
\newcommand{\eTTvec}{\epsilon_{\tau\tau}^\mathrm{iV}}
\newcommand{\eMMu}{\epsilon_{\mu\mu}^{uV}}
\newcommand{\eMMd}{\epsilon_{\mu\mu}^{dV}}

\begin{document}
	
	\title{Complementarity Between Neutrino Neutral and Charged Current Events
		\texorpdfstring{\\}{ }
		in the Search for New Physics}
	\vspace*{-1.5cm}
	\hfill{FERMILAB-PUB-26-0271-T}

	\author{Julia Gehrlein}
	
	\email{julia.gehrlein@colostate.edu}
	\thanks{\orcid{0000-0002-1235-0505}}
	\affiliation{Physics Department, Colorado State University, Fort Collins, CO 80523, USA}

	\author{Jaime Hoefken Zink}
	
	\email{jaime.hoefkenzink@ncbj.gov.pl}
	\thanks{\orcid{0000-0002-4086-2030}}
	\affiliation{National Centre for Nuclear Research, Pasteura 7, Warsaw, PL-02-093, Poland}

	\author{Pedro A.~N.~Machado}
	\email{pmachado@fnal.gov}
	\thanks{\orcid{0000-0002-9118-7354}}
	\affiliation{Theory Division, Fermi National Accelerator Laboratory, Illinois 60510, USA}

	\author{João Paulo Pinheiro}
	\email{joaopaulo.pinheiro@sjtu.edu.cn}
	\thanks{\orcid{0000-0002-6536-2040}}
	\affiliation{State Key Laboratory of Dark Matter Physics, Tsung-Dao Lee Institute \& School of Physics and Astronomy, Shanghai Jiao Tong University, Shanghai 200240, China}
	\affiliation{Key Laboratory for Particle Astrophysics and Cosmology (MOE) \& Shanghai Key Laboratory for Particle Physics and Cosmology, Shanghai Jiao Tong University, Shanghai 200240, China}

	\date{\today}

	\begin{abstract}
		At long-baseline neutrino experiments, neutral-current (NC) events accumulate in large numbers but are seldom exploited for new physics searches.
		We demonstrate their potential using non-standard neutrino interactions (NSI) with quarks as a case study.
		Charged-current (CC) analyses constrain NSI through matter effects on neutrino propagation, which probe almost exclusively the isoscalar combination of up- and down-quark couplings; the orthogonal isovector combination is suppressed by a factor of $\sim$100.
		Because NSI also modify NC cross sections in a flavor-dependent way, NC events become sensitive to oscillations: the far-to-near detector ratio acquires a dependence on the beam's flavor composition that probes both isoscalar and isovector couplings with comparable weight.
		Using existing NOvA data and DUNE projections, we derive the first bounded constraints on isovector NSI from a long-baseline experiment and show that combining CC and NC measurements resolves the individual quark couplings, breaking a degeneracy that persists in either analysis alone.
	\end{abstract}
	
	\maketitle

	\section{Introduction}
	\label{sec:intro}
	
	Neutrino physics has entered a precision era.
	The solar and reactor mixing parameters have been measured with outstanding accuracy~\cite{JUNO:2025gmd, SNO:2025koj, DayaBay:2022eyy, RENO:2025wbu, BOREXINO:2023ygs, Decowski:2016axc}, and measurements of atmospheric mixing parameters continue to be refined~\cite{NOvA:2025tmb, T2K:2025yoy, IceCubeCollaboration:2023wtb, T2K:2024wfn}.
	Several key questions remain open: the octant of $\theta_{23}$, whether CP violation occurs in the leptonic sector, and the neutrino mass ordering~\cite{Esteban:2024eli}.
	Resolving these requires sensitivity to matter effects and precise control of the neutrino energy and baseline, which long-baseline accelerator experiments are designed to provide.
	
	Accelerator long-baseline (LBL) experiments such as NOvA and, in the future, DUNE \cite{DUNE:2020lwj, DUNE:2020ypp, DUNE:2020mra, DUNE:2020txw}, use accelerated protons to produce muon-neutrino beams and deploy a near/far detector pair.
	In a charged-current (CC) interaction, the outgoing charged lepton tags the neutrino flavor and allows kinematic reconstruction of the event.
	Comparing the CC sample at the far and near detectors encodes the oscillation signal, and the CC channel is the primary tool for measuring the atmospheric mixing sector and addressing the open questions above.
	
	The neutral-current (NC) sample also accumulates in large numbers at both detectors.
	In NC interactions, the scattered neutrino escapes undetected, leaving only the hadronic recoil visible; the missing energy makes event reconstruction harder and removes any flavor information.
	Because NC interactions are flavor-universal, any neutrino flavor produces the same signal, and the NC rate carries no oscillation information within the Standard Model.
	The NC event rates are therefore expected to be identical at near and far detectors, up to a geometric acceptance factor.
	Consequently, NC events are treated as background in oscillation analyses.
	
	Long-baseline experiments also probe new physics beyond the standard oscillation framework.
	Non-standard neutrino interactions (NSI) provide a model-independent parametrization of new physics via dimension-six operators coupling neutrinos to charged fermions~\cite{Proceedings:2019qno}.
	The CC sample constrains NSI through their effect on the matter potential, but NSI also modify the NC detection cross section.

	Propagation through matter does not probe all NSI operators individually, but only a specific combination of them.
	Considering NSI that couple only to quarks, and given that the neutron-to-proton ratio in the Earth's crust is approximately unity, the Earth matter potential is sensitive only to the sum of proton and neutron contributions, conveniently characterized as the isoscalar component, invariant under $u \leftrightarrow d$ exchange.
	In contrast, any quantity proportional to the difference between up- and down-quark couplings corresponds to the isovector component, to which propagation is essentially blind.
	Neither increased statistics nor global fits combining several oscillation experiments can access the isovector direction~\cite{Proceedings:2019qno, Farzan:2017xzy}.
	Other probes, including CE$\nu$NS and solar neutrino analyses, face their own limitations that have so far prevented bounded constraints on the isovector combination.
	
	In this work, we show that NC events provide the missing sensitivity.
	The NC cross sections depend on both isoscalar and isovector components with comparable weights, breaking the degeneracy inherent to CC-based analyses.
	Using existing NOvA data and DUNE projections, we derive the first bounded constraints on isovector NSI from a long-baseline experiment, and show that NC constraints on the isoscalar combination are competitive with CC bounds.
	Combining CC and NC measurements resolves the individual up- and down-quark NSI couplings, breaking a degeneracy that persists in either analysis alone.

	\section{Isoscalar and Isovector NSI}
	\label{sec:isosc_isovec_def}
	
	Non-standard neutrino interactions provide a model-independent
	framework to parametrize new physics in the neutrino sector. 
	In the context of neutral-current NSI, coupling neutrinos to matter fermions, 
	the effective Lagrangian reads~\cite{Wolfenstein:1977ue, Farzan:2017xzy, Proceedings:2019qno}
	\begin{equation}
		\mathcal{L}_{\text{NSI}}^{\text{NC}} = -2\sqrt{2}G_F \sum_{f=u,d,e}\sum_{P=L,R} \eps_{\alpha\beta}^{fP} (\bar{\nu}_\alpha \gamma^\mu P_L \nu_\beta)(\bar{f}\gamma_\mu P_P f),
		\label{eq:nsi_lagrangian}
	\end{equation}
	where the interaction has been normalized to the Fermi constant $G_F$, $\eps_{\alpha\beta}^{fP}$ are dimensionless 
	parameters, and $P_{L,R} = (1 \mp \gamma_5)/2$.
	For unpolarized matter, only the
	vector combination
	\begin{equation}
		\eps_{\alpha\beta}^{fV} \equiv
		\eps_{\alpha\beta}^{fL} + \eps_{\alpha\beta}^{fR}
	\end{equation}
	is relevant, while the other combination 
	$\eps_{\alpha\beta}^{fL} - \eps_{\alpha\beta}^{fR}$
	is defined as axial NSI, $\eps_{\alpha\beta}^{fA}$.
	Axial NSI do not contribute to coherent forward scattering and are therefore
	absent from the matter potential~\cite{Farzan:2017xzy};
	their effects on NC cross sections have been studied in
	Refs.~\cite{Abbaslu:2023vqk,Abbaslu:2024hep,Abbaslu:2025fwt} and we do not consider them further in this work.
	The physical effects of quark-coupled NSI can be
	decomposed into isoscalar and isovector components,
	\begin{align}
		\eps_{\alpha\beta}^\mathrm{iS(V,A)}
		&\equiv \eps_{\alpha\beta}^{uV,A} + \eps_{\alpha\beta}^{dV,A}
		\quad \text{(isoscalar)}\,,\\
		\eps_{\alpha\beta}^\mathrm{iV(V,A)}
		&\equiv \eps_{\alpha\beta}^{uV,A} - \eps_{\alpha\beta}^{dV,A}
		\quad \text{(isovector)}\,,
		\label{eq:isoscalar_isovector}
	\end{align}
	where the naming reflects their transformation under isospin symmetry~\cite{Gell-Mann:1964ewy}.
	
	We note that vector NSI arise naturally in UV-complete extensions of the Standard
	Model, including models with additional $Z'$ gauge bosons, leptoquark models,
	and models with scalar
	mediators~\cite{Babu:2017olk,Proceedings:2019qno,Babu:2019mfe,Coloma:2020gfv,Bernal:2022qba,DeRomeri:2023cjt}.
	In many of these constructions the couplings to up and down quarks are
	generically different, yielding a naturally nonzero isovector component
	$\eps^\mathrm{iV}$, precisely the direction to which
	propagation-based oscillation experiments are blind.
	
	\section{Effect of NSI in CC and NC events }
	\label{sec:nsi_effects}
	\subsection{CC Events: Sensitivity Through Propagation}
	\label{sec:cc_theory}
	At long-baseline experiments, charged-current events are used to identify the neutrino flavor after propagation through matter and extract the oscillation probability. This probability depends on the effective Hamiltonian in the flavor basis,
	\begin{equation}
		H = \frac{1}{2E} U \begin{pmatrix} 0 & & \\ & \Delta m_{21}^2 & \\ & & \Delta m_{31}^2 \end{pmatrix} U^\dagger + V_{\text{mat}},
	\end{equation}
	where $U$ is the PMNS mixing matrix~\cite{Pontecorvo:1957cp,Maki:1962mu}, $E$ is the neutrino energy, and $\Delta m_{ij}^2 = m_i^2 - m_j^2$, where $m_i$ for $i=1,2,3$ are the neutrino masses.
	
	The matter potential matrix in the presence of NSI is
	\begin{equation}
		V_{\text{mat}} = \sqrt{2}G_F N_e \begin{pmatrix}
			1 + \eps_{ee} & \eps_{e\mu} & \eps_{e\tau} \\
			\eps_{e\mu}^{*} & \eps_{\mu\mu} & \eps_{\mu\tau} \\
			\eps_{e\tau}^{*} & \eps_{\mu\tau}^{*} & \eps_{\tau\tau}
		\end{pmatrix},
	\end{equation}
	where the effective propagation NSI are written as
	\begin{equation}
		\eps_{\alpha\beta} = \eps_{\alpha\beta}^{eV} + \frac{N_u}{N_e}\eps_{\alpha\beta}^{uV} + \frac{N_d}{N_e}\eps_{\alpha\beta}^{dV}~,
		\label{eq:eps_prop}
	\end{equation}
	and setting all of them to zero recovers the standard matter potential.
	For electrically neutral matter ($N_e = N_p$), where $N_e$, $N_p$, $N_n$ are the electron, proton, and neutron number densities, the valence quark content of the nucleons gives $N_u/N_e = 2 + Y_n$ and $N_d/N_e = 1 + 2Y_n$, where $Y_n \equiv N_n/N_e$ is the neutron-to-electron ratio. Substituting into Eq.~\eqref{eq:eps_prop} and applying the isospin decomposition we obtain
	\begin{align}
		\eps_{\alpha\beta} &= \eps^{eV}_{\alpha\beta} + (2 + Y_n)\eps^{uV}_{\alpha\beta} + (1 + 2Y_n)\eps^{dV}_{\alpha\beta} \nonumber\\
		&= \eps^{eV}_{\alpha\beta} + \underbrace{3\frac{(1 + Y_n)}{2}\eps^\mathrm{iS}_{\alpha\beta}}_{\text{isoscalar}} + \underbrace{\frac{1 - Y_n}{2}\eps^\mathrm{iV}_{\alpha\beta}}_{\text{isovector}}~.
		\label{eq:eps_prop_isospin}
	\end{align}
	We henceforth set electron NSI to zero and restrict to diagonal flavor indices, $\alpha = \beta \in \{\mu, \tau\}$.
	For the Earth's crust and mantle, $Y_n \approx 1.05$, and thus
	\begin{equation}
		\boxed{\eps^{\oplus}_{\alpha\alpha} \approx 3.08\,\eps^\mathrm{iS}_{\alpha\alpha} - 0.025\,\eps^\mathrm{iV}_{\alpha\alpha}}\,.
		\label{eq:eps_prop_earth}
	\end{equation}
	The isovector contribution is suppressed by a factor of $\sim$100 relative to the isoscalar one, rendering CC oscillation measurements essentially blind to this component.
	
	\subsection{NC Events: Sensitivity Through Cross Sections}
	\label{sec:nc_theory}
	
\begin{figure}
		\centering
		\includegraphics[width=0.9\linewidth]{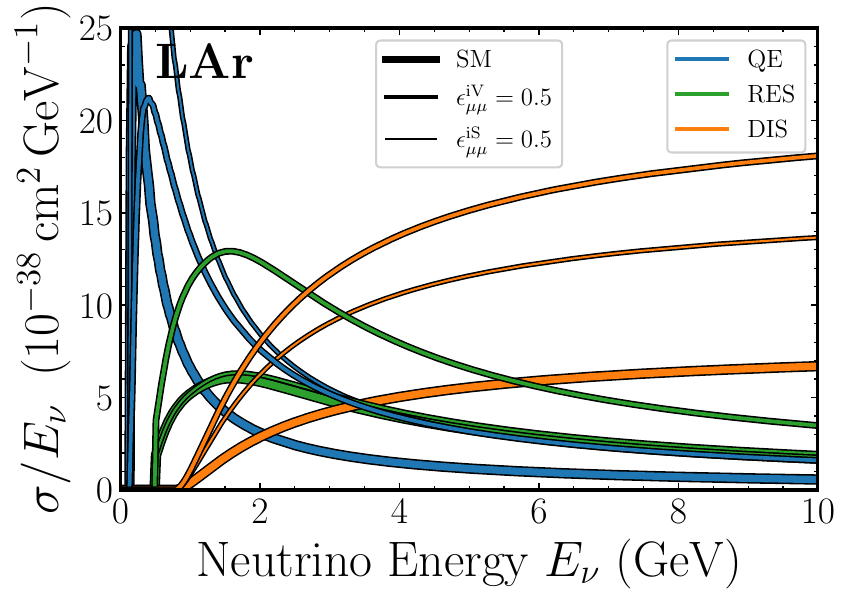}
		\includegraphics[width=0.9\linewidth]{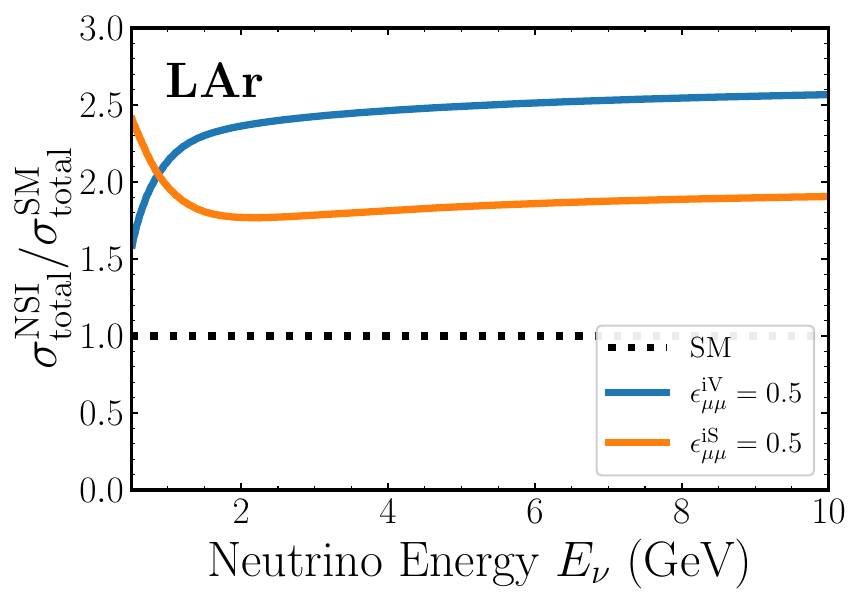}
		\caption{Neutral-current neutrino-argon cross sections.
			Upper panel: individual channel contributions (QE in blue, RES in green, DIS in orange) for the SM (thick lines), isovector NSI with $\eps^\mathrm{iV}_{\mu\mu}=0.5$ (medium), and isoscalar NSI with $\eps^\mathrm{iS}_{\mu\mu}=0.5$ (thin).
			Lower panel: ratio of the total NSI to SM cross section (isovector in blue, isoscalar in orange).}
		\label{fig:xsec}
	\end{figure}

	NC events provide complementary sensitivity to NSI through modifications 
	of the scattering cross section. In the energy regime relevant for DUNE 
	and NOvA (0.5--10~GeV), three mechanisms dominate: quasi-elastic scattering 
	(QE) $\nu + N \to \nu + N$, resonant single pion production (RES) 
	$\nu + N \to \nu + N' + \pi$, and deep inelastic scattering (DIS) 
	$\nu + N \to \nu + X$, where $N$, $N'$ are nucleons and $X$ 
	represents hadronic activity. The cross section for argon\footnote{NC detection involves finite momentum transfer $q^2 \sim (0.1\text{--}1~\text{GeV})^2$. For the EFT description in Eq.~\eqref{eq:nsi_lagrangian} to remain valid, the mediator must satisfy $M \gtrsim \mathcal{O}(1~\text{GeV})$~\cite{Farzan:2017xzy, Coloma:2020gfv}. 
	} is shown in Fig.~\ref{fig:xsec}.

	With NSI, the effective nucleon vector couplings become
	\begin{align}
		\tilde{g}_{V,\alpha}^p &= g_V^p + \frac{3\eps^\mathrm{iS}_{\alpha\alpha} + \eps^\mathrm{iV}_{\alpha\alpha}}{2}, \\
		\tilde{g}_{V,\alpha}^n &= g_V^n + \frac{3\eps^\mathrm{iS}_{\alpha\alpha} - \eps^\mathrm{iV}_{\alpha\alpha}}{2},
	\end{align}
	where $g_V^p = 1/2 - 2\sw \approx 0.04$ and $g_V^n = -1/2$ are the SM values, with $\sin^2\theta_W$ being the weak mixing angle.
	Unlike in propagation, the NC cross sections depend on both the isoscalar
	and isovector components with comparable weights.
	The isovector combination $\eps^\mathrm{iV}$ enters the pion-production transition amplitudes for the $\Delta(1232)$ and $N(1520)$ resonances, which dominate the RES cross section.
	For details of the cross-section implementation, see the companion paper~\cite{companion}.
	
	CE$\nu$NS offers an independent probe of the same NSI couplings. In the presence of NSI, the coherent weak charge of a nucleus for neutrino flavor $\alpha$ is
	\begin{equation}
		Q_{w\alpha} = Zg_V^p + Ng_V^n + \frac{3(Z+N)}{2}\eps^\mathrm{iS}_{\alpha\alpha} + \frac{Z-N}{2}\eps^\mathrm{iV}_{\alpha\alpha}~,
		\label{eq:Qw}
	\end{equation}
	and the cross section scales as $Q_{w\alpha}^2$. For targets with $Z \approx N$, the isovector coefficient $(Z-N)/2$ is suppressed and CE$\nu$NS measures only the isoscalar direction; combining results from nuclei with different $Z/N$ ratios could in principle disentangle the two couplings, but not given current precision~\cite{Liao:2024qoe}. CE$\nu$NS experiments are also insensitive to $\eps_{\tau\tau}^{fV}$, as no $\nu_\tau$ source is available~\cite{COHERENT:2017ipa, COHERENT:2020iec, COHERENT:2021xmm, COHERENT:2020ybo}.
	
	Solar neutrino experiments~\cite{SNO:2009uok, SNO:2011hxd} are not subject to the $Y_n \approx 1$ suppression: in the Sun $Y_n$ ranges from $\sim 0.5$ at the core to $\sim 0.14$ at the surface. 
	However, the oscillation Hamiltonian constrains only the differences $(\eps_{\alpha\alpha} - \eps_{\beta\beta})$, leaving the individual couplings undetermined. 
	In contrast, NC events at long-baseline experiments probe the absolute values of $\eps^\mathrm{iS}$ and $\eps^\mathrm{iV}$ through their effect on the cross section, and the large $\nu_\tau$ component at the far detector provides additional sensitivity to tau-sector NSI.
	
	\begin{figure}[!ht]
		\centering
		\includegraphics[width=0.9\linewidth]{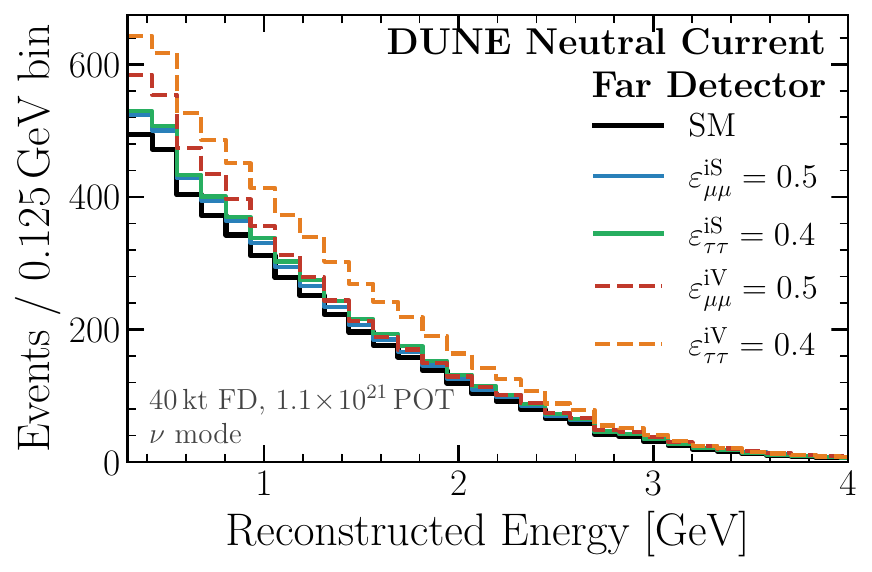}
		\includegraphics[width=0.9\linewidth]{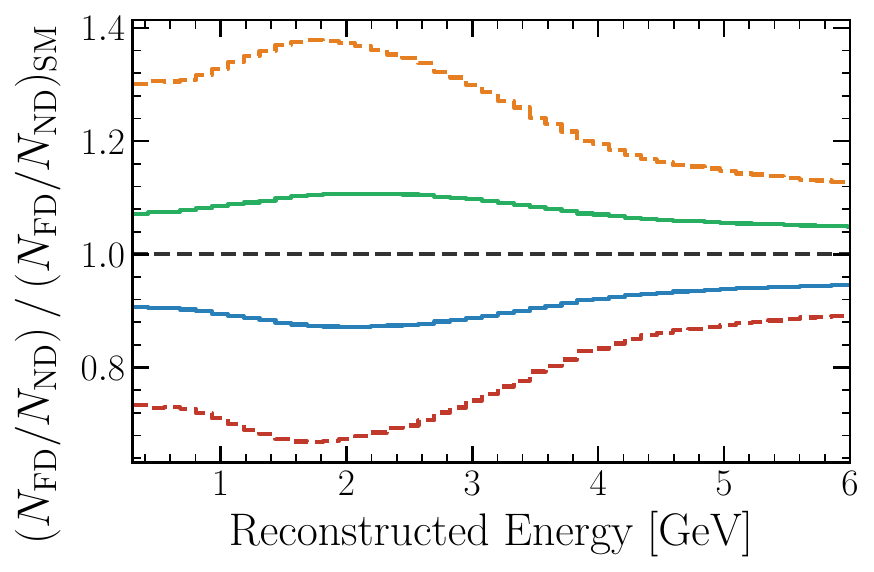}
		\caption{Effect of isoscalar and isovector NSIs on the event distributions
			at DUNE. (a)~Distribution of NC events at the DUNE FD as a function of
			reconstructed energy for different NSI benchmark values: $\eMMsc = 0.5$,
			$\eTTsc = 0.4$, $\eMMvec = 0.5$, and $\eTTvec = 0.4$. (b)~Ratio of FD to ND
			event rates, normalized to the SM expectation, as a function of reconstructed energy for the same benchmark points.
		}
		\label{fig:nova_dune_dist}
	\end{figure}

	\section{Results}
	\label{sec:results}
	
	We analyze existing NOvA data, using both NC~\cite{NOvA:2024imi}
	and CC~\cite{NOvA:nu24} event samples, alongside projected DUNE
	sensitivities based on simulated pseudo-data assuming no NSI. The
	DUNE mock data configuration assumes a total exposure of
	$1.1\times 10^{21}$~POT/year with equal running times of 6.5~years
	in forward and reverse horn current configurations.
	The charged-current analysis probes NSI through their effect on neutrino
	oscillation probabilities during propagation through Earth matter. We analyze
	both appearance ($\nu_\mu \to \nu_e$) and disappearance ($\nu_\mu \to \nu_\mu$)
	channels for neutrinos and antineutrinos, marginalizing over the
	atmospheric mixing parameters. To simulate CC events
	at DUNE we use the GLoBES~\cite{Huber:2007ji} implementation of Ref.~\cite{DUNE:2021cuw}; for NOvA we use
	a dedicated simulation framework developed by one of the authors~\cite{Pinheiro:2025lqk}.
	For the NC analysis, we employ a combined fit of near detector (ND)
	and far detector (FD) data, exploiting the FD/ND ratio to cancel
	common systematic uncertainties, following the approach developed
	in Ref.~\cite{Gehrlein:2024vwz}.
	Details of the analysis procedure are given in Appendix~\ref{sec:details_analysis}.
	
	\afterpage{%
		\begin{figure*}[tp]
			\centering
			\includegraphics[width=\linewidth]{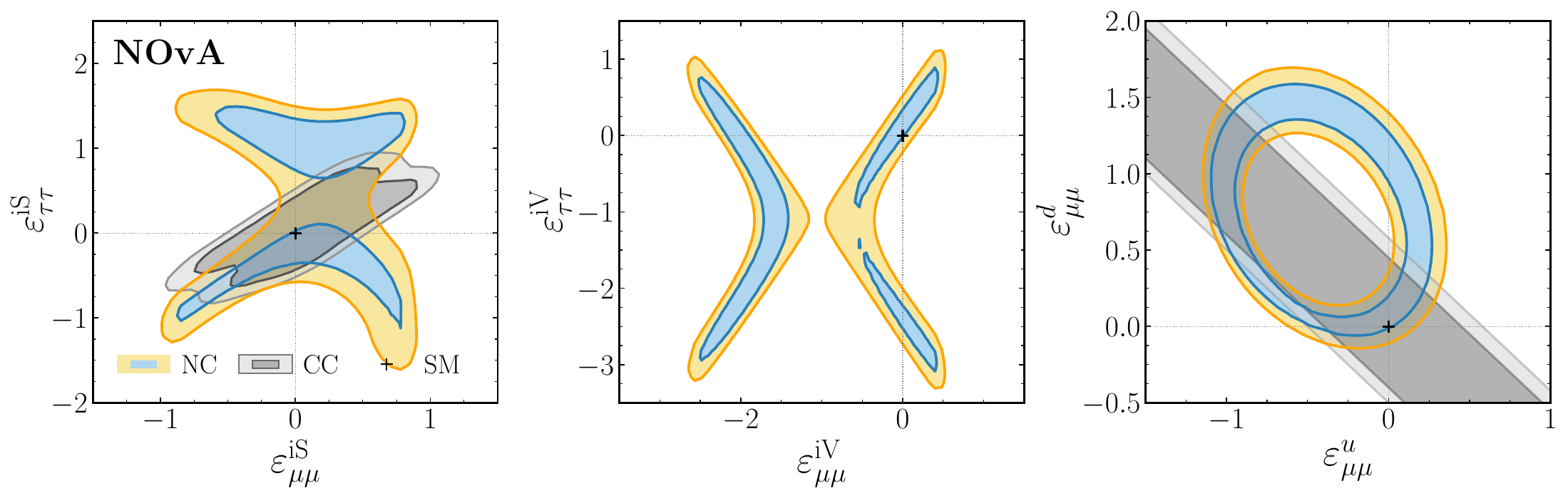}
			\caption{Constraints on NSI parameters from NOvA at 1$\sigma$ and 2$\sigma$ CL,
				showing NC (blue, light orange) and CC (dark gray, light gray) analyses.
				\textit{Left:} isoscalar $\eMMsc$--$\eTTsc$ plane.
				\textit{Center:} isovector $\eMMvec$--$\eTTvec$ plane (NC only, as CC is insensitive).
				\textit{Right:} $\eMMu$--$\eMMd$ plane, where the diagonal CC band reflects sensitivity to only $\eps^{uV} + \eps^{dV}$ and NC breaks this degeneracy.}
			\label{fig:contours_nova}
		\end{figure*}
	}
	
	We first examine how NSI affect event distributions, focusing on DUNE for concreteness.
	Figure~\ref{fig:nova_dune_dist}, top panel, presents the expected event distribution 
	in the DUNE FD as a function of the reconstructed energy for four representative 
	values of NSI parameters: $\eMMsc = 0.5$, $\eTTsc = 0.4$, $\eMMvec = 0.5$, and $\eTTvec = 0.4$. 
	As expected from the cross-section formulas discussed in Ref.~\cite{companion}, the number of NC events grows quadratically with the NSI parameter. 
	However, the effect of isovector couplings is stronger 
	than that of isoscalar ones, since RES scattering is predominantly affected by isovector couplings.

	Since our NC analysis relies on the FD/ND ratio, the bottom panel of Fig.~\ref{fig:nova_dune_dist} shows this double ratio, normalized to the SM expectation, for the same benchmark points.
	The $\tau$-sector NSI ($\eTTsc$, $\eTTvec$) increase the ratio above unity because oscillations convert part of the $\nu_\mu$ beam into $\nu_\tau$ at the FD, while the ND contains essentially no $\nu_\tau$; any enhancement of $\nu_\tau$ NC scattering therefore affects only the FD.
	Conversely, $\mu$-sector NSI ($\eMMsc$, $\eMMvec$) suppress the ratio below unity because the ND, being dominated by $\nu_\mu$, receives a proportionally larger cross-section enhancement than the FD, where part of the $\nu_\mu$ flux has oscillated away.
	In both sectors, the isovector benchmarks produce a larger deviation from unity than the isoscalar ones at the same parameter value, consistent with the larger cross-section modifications seen in the top panel.
	
	For isoscalar NSI, an additional effect partially cancels the sensitivity: the isoscalar component enters the matter potential, see Eq.~\eqref{eq:eps_prop_earth}, and modifies the $\nu_\mu$ survival probability, pushing the FD flavor composition back toward the ND composition and the FD/ND ratio toward unity.
	Isovector NSI do not share this cancellation, as they do not change matter effects significantly.

	\afterpage{%
		\begin{figure*}[tp]
			\centering
			\includegraphics[width=\linewidth]{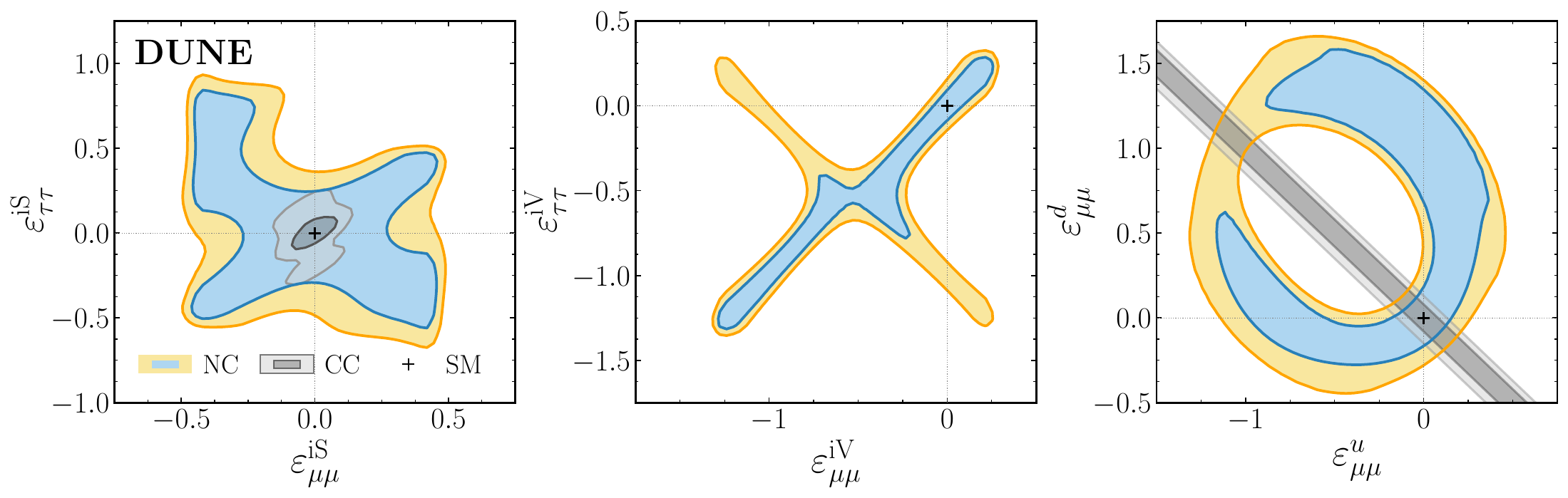}
			\caption{Projected constraints on NSI parameters from DUNE at 1$\sigma$ and 2$\sigma$ CL (same layout as Fig.~\ref{fig:contours_nova}).
				The stronger bounds and multi-island CC structure reflect DUNE's longer baseline and larger matter effects.}
			\label{fig:contours_dune}
		\end{figure*}
	}

	The left panels of Figs.~\ref{fig:contours_nova} and~\ref{fig:contours_dune} show the constraints in the $\eMMsc$--$\eTTsc$ plane from CC and NC events at NOvA (observed) and DUNE (projected).
	For NOvA CC (Fig.~\ref{fig:contours_nova}, left), the allowed regions are larger due to the shorter baseline,
	detector geometry, and smaller neutrino flux. Only two connected regions
	appear, a consequence of using real data, where the central island merges
	with the surrounding structure. At NOvA's shorter baseline, CC and NC
	constraints on isoscalar NSI are comparable, and the complementarity
	between these two data samples improves the overall sensitivity.
	
	For DUNE, the CC contours (dark and light gray) exhibit a characteristic multi-island
	structure, similar to the results in Ref.~\cite{DUNE:2021cuw}, arising
	from degeneracies between oscillation parameters and NSI. The islands
	are centered around the SM point. Due to DUNE's long baseline and
	correspondingly large matter effects, the CC constraints on isoscalar
	NSI are significantly stronger than those from NC events.
	
	The NC contours (blue and light orange) show similar shapes for both
	experiments. This similarity arises because isoscalar NSI enter the
	cross section with comparable weights across QE, RES, and DIS processes,
	effectively acting as an overall rescaling factor. The resulting contours
	closely resemble those obtained with an effective NSI parametrization
	in Ref.~\cite{Gehrlein:2024vwz}. For large NSI values, the FD/ND ratio
	approaches unity; however, the absolute event rate becomes large, and
	these regions are disfavored by our BSM penalty term in Eq.~\eqref{eq:bsm_pull}.
	
	The NC bounds from DUNE are stronger than those from NOvA primarily
	due to smaller assumed systematic uncertainties.
	Additionally, the DUNE contours are more symmetric since they 
	represent a sensitivity study assuming no-NSI pseudo-data, whereas the 
	NOvA contours reflect actual measurements.

	As discussed before, CC events are essentially blind to isovector NSI due to the approximate equality $N_n \approx N_e$ in the Earth's crust and mantle, leaving NC cross sections as the primary probe of this direction in parameter space.
	The center panels of Figs.~\ref{fig:contours_nova} and~\ref{fig:contours_dune} present the constraints in the
	$\eMMvec$--$\eTTvec$ plane from NC events alone; CC analyses have
	no sensitivity in this direction. These constraints represent genuinely
	new information inaccessible to oscillation-based searches.
	
	The structure of the isovector contours depends on the dominant
	cross-section mechanism at each experiment's characteristic energy.
	At NOvA's peak energy of $\sim$2~GeV, resonance production dominates,
	leading to a degeneracy that manifests as two disconnected
	allowed regions in Fig.~\ref{fig:contours_nova} (center). At DUNE's higher energies,
	DIS becomes more significant, and the contours remain symmetric but
	the gap between the two regions closes, connecting them into a single
	allowed region in Fig.~\ref{fig:contours_dune} (center).

	The right panels of Figs.~\ref{fig:contours_nova} and~\ref{fig:contours_dune} display
	constraints in the $(\eMMu, \eMMd)$ plane, revealing the geometrical origin of the CC--NC complementarity.
	
	The CC constraints appear as diagonal bands oriented along lines of constant $\eps^{uV} + \eps^{dV}$. This reflects the
	exclusive sensitivity of matter-induced oscillations to the
	isoscalar combination: negative $\eps^{uV}$ can be compensated by positive 
	$\eps^{dV}$ (and vice versa) without affecting the propagation 
	Hamiltonian. 
	This degeneracy persists regardless of 
	experimental precision; it is a geometric blind direction, not
	a statistical limitation.
	
	The NC constraints from NOvA (DUNE), shown as elliptical regions,
	break this degeneracy by probing $\eps^{uV}$ and $\eps^{dV}$ 
	independently. The combination of CC and NC data at long-baseline experiments provides a unique path to individual constraints on $\eps^{uV}$ and $\eps^{dV}$
	from a single experiment, accessing parameter space inaccessible to
	either analysis alone or to current CE$\nu$NS measurements.
	
	Table~\ref{tab:nsi_limits} summarizes the combined CC+NC constraints on the diagonal NSI parameters, for which the two analyses provide complementary information. For off-diagonal parameters, the CC bounds dominate and the combination is less informative.
	As anticipated, DUNE is expected to improve upon the current NOvA constraints by a factor of $2$--$3$.
	
	\begin{table}[b]
		\centering
		\caption{Current and projected constraints on NSI parameters from the combined CC+NC analysis at NOvA
			and DUNE
			, respectively.
			Round and square brackets denote 90\% and 95\% confidence levels assuming Wilks' theorem.}
		\label{tab:nsi_limits}
		\begin{tabular}{lcc}
			\hline\hline
			NSI & NOvA & DUNE \\
			\hline
			$\varepsilon_{\tau\tau}^{uV}$ & \makecell{$(-0.22, 0.21)$\\$[-0.25, 0.25]$} & \makecell{$(-0.063, 0.054)$\\$[-0.080, 0.066]$} \\[1ex]
			$\varepsilon_{\tau\tau}^{dV}$ & \makecell{$(-0.17, 0.13)$\\$[-0.21, 0.19]$} & \makecell{$(-0.064, 0.056)$\\$[-0.080, 0.070]$} \\[1ex]
			$\varepsilon_{\mu\mu}^{uV}$   & \makecell{$(-0.18, 0.10)$\\$[-0.25, 0.14]$} & \makecell{$(-0.060, 0.069)$\\$[-0.074, 0.090]$} \\[1ex]
			$\varepsilon_{\mu\mu}^{dV}$   & \makecell{$(-0.10, 0.17)$\\$[-0.14, 0.21]$} & \makecell{$(-0.059, 0.072)$\\$[-0.073, 0.099]$} \\
			\hline\hline
		\end{tabular}
	\end{table}

	\section{Conclusions}
	\label{sec:conclusion}
	Neutral-current events at long-baseline experiments are routinely treated as background, yet they carry independent sensitivity to non-standard neutrino interactions through cross-section modifications~\cite{Gehrlein:2024vwz}.
	We have shown that CC and NC samples probe orthogonal directions in the NSI parameter space: CC constrains the isoscalar combination $\eps^{uV} + \eps^{dV}$ through the matter potential, while NC cross sections depend on both isoscalar and isovector components with comparable weights, accessing a direction invisible to oscillation-based searches.

	Using existing NOvA data, we have derived the first bounded constraints on isovector NSI from a long-baseline experiment.
	The combined CC+NC analysis constrains the individual quark couplings $\eps_{\mu\mu}^{uV}$, $\eps_{\mu\mu}^{dV}$, $\eps_{\tau\tau}^{uV}$, and $\eps_{\tau\tau}^{dV}$ at the $\ord{0.2}$ level at 90\% CL.
	Projected DUNE sensitivities improve these bounds to $\ord{0.07}$, a factor of 2--3 over current NOvA constraints.
	In the $(\eps^{uV}, \eps^{dV})$ plane, the CC-allowed region forms a diagonal band along the isoscalar direction, while the NC contours form a closed region that intersects this band, breaking the degeneracy.
	
	The combination of CC and NC measurements at a single long-baseline experiment provides complete coverage of the $(\eps^{uV}, \eps^{dV})$ plane for both $\mu\mu$ and $\tau\tau$ NSI, resolving a degeneracy that persists in oscillation-only analyses, solar neutrino data, and current CE$\nu$NS measurements.
	Our analysis employs a conservative framework based on FD/ND event ratios; a full experimental analysis with detector-level simulations and optimized NC event selection could strengthen these constraints.
	We encourage the NOvA and DUNE collaborations to incorporate NC event samples into their searches for non-standard neutrino interactions.

	\section*{Acknowledgments}
	We thank Enrique Fernandez-Martinez for kindly providing us with the migration matrices for DUNE NC events from Ref.~\cite{DeRomeri:2016qwo} and Maria Concepcion Gonzalez-Garcia for useful discussions and comments.
	JG acknowledges support by
	the U.~S.~ Department of Energy Office of Science under award number DE-SC0025448.
	P.M. is supported by Fermi Forward Discovery Group, LLC under Contract No. 89243024CSC000002 with the U.S. Department of Energy, Office of Science, Office of High Energy Physics. 
	JPP is supported by the National Natural Science Foundation of China (12425506 and 12375101). This work is also supported by State Key Laboratory of Dark Matter Physics. J.~H.~Z. is supported by the National Science Centre, Poland (research grant No. 2021/42/E/ST2/00031).

	\appendix
	\section{Details on the analysis}
	\label{sec:details_analysis}

	\subsection{NC events}
	
	Neutrino cross-section modeling at LBL experiments is known to
	exhibit significant discrepancies with data~\cite{NOvA:2020rbg, MicroBooNE:2021ccs}:
	measured neutrino spectra at the ND systematically deviate from
	baseline generator predictions, requiring experiment-specific
	tuning that propagates into the unfolding and migration procedures~\cite{Coyle:2025xjk}.
	Since the resulting cross-section model depends on the particular
	tuning adopted by each experiment, absolute event rate predictions
	carry large model-dependent uncertainties.
	To mitigate potential cross section mis-modeling and tuning dependence, we construct our test statistic using near-to-far detector ratios rather than from the individual detector rates.
	In this
	ratio, flux and cross-section uncertainties largely cancel,
	yielding a conservative but robust framework for constraining
	NSI through NC events.
	
	The expected event rate in true energy bins is computed as
	\begin{equation}
		N_i^{\text{true}} = C \int_{E_i}^{E_{i+1}} \!\!dE_\nu \, \frac{d\Phi}{dE_\nu} \,
		\sum_N\text{Tr}[\rho(L) \cdot \sigma_N(E_\nu)]~,
	\end{equation}
	where $C$ encodes detector-specific normalization factors,
	$d\Phi/dE_\nu$ is the differential neutrino flux, $\rho(L)$ is the flavor density matrix
	after propagation over baseline $L$, and $\sigma_N(E_\nu)$ is the weighted generalized cross-section
	matrix in flavor space~\cite{Coloma:2022umy, Amaral:2023tbs}, with $N$  the type of nucleon.
	The trace runs over flavor indices.
	The per-nucleon generalized cross section $\sigma_{\alpha\beta}^{N}$ receives
	contributions from quasi-elastic (QE), resonance (RES), and deep
	inelastic scattering (DIS):
	\begin{equation}
		{\sigma_N}_{\alpha\beta} = N_N(\sigma_{\alpha\beta}^{N,\text{QE}} +
		\sigma_{\alpha\beta}^{N,\text{RES}} + \sigma_{\alpha\beta}^{N,\text{DIS}}),
	\end{equation}
	where $N_N$ is the number of nucleons $N$ in each target. The BSM cross sections are obtained by modifying the SM vector couplings $g_V^{p,n}$ as described in Sec.~\ref{sec:nc_theory}, and reweighting the SM NuWro prediction~\cite{Golan:2012rfa} channel by channel; the full procedure will be detailed in Ref.~\cite{companion}.
	
	The predicted event spectrum in reconstructed energy is obtained
	from the true energy distribution through a two-step procedure.
	First, we apply the migration matrix $M_{ij}$, which maps events
	from true energy bin $i$ to reconstructed energy bin $j$, encoding
	the detector response and energy reconstruction effects. For NOvA,
	the migration matrix is constructed from Monte Carlo simulations
	using the NuWro generator~\cite{Golan:2012rfa}, following the
	procedure described in Ref.~\cite{Gehrlein:2024vwz}; for DUNE,
	we adopt the migration matrix from Ref.~\cite{DeRomeri:2016qwo}.
	The resulting spectrum is then weighted by the selection efficiency
	$R_j$, which accounts for the fraction of NC events in
	reconstructed bin $j$ that survive the event selection criteria
	and are correctly identified as neutral-current interactions,
	\begin{equation}
		N_j = R_j \sum_i M_{ij} N_i^{\text{true}}.
	\end{equation}
	
	With the reconstructed events we
	build a $\chi^2$ test statistic that simultaneously fits ND and FD data
	with correlated systematic uncertainties,
	\begin{align}
		\chi^2 &= \sum_j^\text{bins} \frac{\left[(1+x_j)N_j - D_j\right]^2}{D_j} + \frac{\left[(1+x_j+y)N'_j - D'_j\right]^2}{D'_j} \nonumber\\
		&\quad+ \left(\frac{y}{\sigma_y}\right)^2 + \chi^2_{\rm NDtot}+ \chi^2_{\rm FDtot},
		\label{eq:chisq_NC}
	\end{align}
	where $D_j$ denotes the data (or mock data) in bin $j$.
	The unprimed and primed quantities denote near and far detector events, respectively.
	The $\chi^2$ is always minimized over the pull parameters $x_j$ and $y$.
	The pull parameters $x_j$ are unconstrained bin-by-bin shape uncertainties, fully correlated between near and far detector, while $y$ is a near-to-far uncorrelated overall normalization factor.
	We use $\sigma_y = 0.15$ for
	NOvA and $\sigma_y = 0.10$ for DUNE, consistent with the NC uncertainties
	in Refs.~\cite{NOvA:2024imi,DUNE:2021cuw}.
	
	Additionally, we include penalty terms informed by existing NC cross-section measurements on argon~\cite{MicroBooNE:2022zhr} and mineral oil~\cite{MiniBooNE:2009dxl}, which agree with SM predictions within 10--30\%.
	These priors constrain the total event rate at each detector and are implemented as
	\begin{equation}
		\chi^2_{\rm NDtot} = \begin{cases}
			\left(\dfrac{N/D - 1.5}{0.1}\right)^2
			& \text{if } N/D > 1.5 \\[8pt]
			0 & \text{otherwise,}
		\end{cases}
		\label{eq:bsm_pull}
	\end{equation}
	where $N$ and $D$ are the total number of predicted events and data in the near detector.
	The prior for the far detector is the same, but for far detector events.
	These terms disfavor BSM scenarios in which the predicted event
	rate exceeds 150\% of the data at either detector and prevent runaway minima for very large NSI.
	
	To simulate NC events at NOvA, we follow the procedure from Ref.~\cite{Gehrlein:2024vwz}, 
	simultaneously fitting ND and FD NC data from Ref.~\cite{NOvA:2024imi} using the migration matrix from the appendix of Ref.~\cite{Gehrlein:2024vwz}. This data consists of $13.6\times 10^{20}$ POT in the  neutrino mode. 
	The target material is CH$_2$, and we use 14 logarithmically-spaced energy bins. 
	The NOvA flux peaks around 2~GeV where resonance production dominates 
	the cross section.

	
	To simulate NC events at DUNE, we follow the simulation framework from Ref.~\cite{DUNE:2021cuw}. We adopt the Technical Design 
	Report (TDR) configuration~\cite{DUNE:2020ypp} with a 40~kton LArTPC far 
	detector and  a ND LAr fiducial volume of 67~t. 
	The target is 
	argon (18 protons, 22 neutrons per nucleus), and we use 125~MeV linear 
	energy bins from 0.3 to 8.0~GeV. The DUNE flux is broad, with QE, RES, 
	and DIS contributions varying across the energy range.
	The setup uses a 120~GeV proton beam with 1.2~MW beam power, 
	resulting in $1.1 \times 10^{21}$~POT/year. We assume equal running time 
	of 6.5 years in forward and reverse horn configurations (neutrino and 
	antineutrino modes), yielding a total exposure of 336~kton$\cdot$MW$\cdot$year.
	We use the migration matrix for NC events at DUNE from \cite{DeRomeri:2016qwo}, see Ref.~\cite{Furmanski:2020smg} for a recent work on neutrino energy estimation in NC events in LAr detectors.

	In both experiments we fix the oscillation parameters to $\theta_{23}=49.6^\circ, \Delta m_{31}^2=2.5\times 10^{-3}~\text{eV}^2$, the best-fit oscillation parameters from \cite{NOvA:2024imi}.
	
	\subsection{CC events}
	
	For DUNE, we perform the experimental simulations for CC events using the GLoBES 
	package~\cite{DUNE:2021cuw}. 
	We use a line-averaged constant Earth matter density of 
	$\rho = 2.848$~g/cm$^3$ following the PREM profile~\cite{Dziewonski:1981xy}. 
	The standard oscillation parameters are fixed to benchmark values consistent 
	with Nufit 6.1 and JUNO first data~\cite{Esteban:2024eli,JUNO:2025gmd}: $\sin^2\theta_{12} = 0.309$, 
	$\sin^2\theta_{13} = 0.022$, and $\Delta m^2_{21} = 7.5 \times 10^{-5}$~eV$^2$. 
	The atmospheric parameters $\Delta m^2_{31}$, $\theta_{23}$, and $\delta_\mathrm{CP}$ 
	are allowed to vary freely in the fit, as their precise determination is among 
	DUNE's primary physics goals. We adopt a 3\% uncertainty on $\theta_{13}$, 
	reflecting constraints from reactor experiments~\cite{DayaBay:2022orm}. 
	Normal mass ordering is assumed throughout.
	The $\chi^2$ is minimized over the standard oscillation to obtain confidence level regions. For details on the NOvA implementation see Ref.~\cite{Pinheiro:2025lqk}.
	
	\bibliography{main}

\end{document}